\documentstyle[12pt,psfig]{article}
\newcommand{\beq}{\begin{equation}}
\newcommand{\eeq}{\end{equation}}

\begin{document}
\begin{center}
\Large
{\bf Chaotic oscillations in finite quantum systems: \\
trapped Bose-Einstein condensates}
\footnote{Short Report to the 4th International Summer School/Conference 
``Let's Face Chaos through Nonlinear Dynamics'', 27 June - 11 July, 
University of Maribor, Slovenia.} 
\normalsize

{\bf Luca Salasnich}\footnote{E-mail: salasnich@mi.infm.it} 
  
{\em Istituto Nazionale per la Fisica della Materia, Unit\`a di Milano, \\
Dipartimento di Fisica, Universit\`a di Milano, \\
Via Celoria 16, 20133 Milano, Italy}
\end{center}                            

\normalsize
\noindent

\begin{center}
{\bf Abstract}
\end{center}
We discuss the recently achieved 
Bose-Einstein condensation for alkali-metal atoms in magnetic traps. 
The theoretically predicted low-energy collective oscillations 
of the condensate have been experimentally confirmed by laser 
imaging techniques. We show by using Poincar\`e sections 
that at higher energies non-linear 
effects appear and oscillations become chaotic.  

\section{Introduction}
\par 
Finite many-body quantum systems, both fermionic and bosonic, 
exhibit collective oscillations, which are nowadays experimentally 
detected with sophisticated devices. In many cases, 
these collective oscillations can become strongly aperiodic 
and also chaotic [1-3]. 
\par
In this short report, we analyze 
the collective modes of a trapped Bose condensate of alkali atoms. 
The Bose-Einstein condensation (BEC) 
is the macroscopic occupation of the ground-state of the system 
of bosons. From 1995 we have experimental results interpreted as an 
evidence of BEC in dilute vapors of confined alkali-metal atoms 
($^{87}$Rb, $^{23}$Na and $^7$Li) [4-6]. 
The experiments with alkali-metal atoms generally consist 
of a laser cooling and confinement in an external potential 
(a magnetic or magneto-optical trap) and 
an evaporative cooling (temperature of the order of $100$ nK) [4-6]. 
Nowadays a dozen of experimental groups have achieved the BEC by using 
different geometries of the confining trap and atomic species. 
\par 
The dynamics of the Bose condensate 
can be accurately described by the Gross-Pitaevskii equation [7] 
of mean-field approximation. 
The theoretically predicted low-energy collective oscillations 
of the condensate have been experimentally confirmed by laser 
imaging techniques [8]. In this short report 
we show that at higher energies non-linear 
effects appear and eventually the collective 
oscillations become chaotic. 

\section{Mean-field theory of BEC}
\par
The many-body problem of $N$ identical bosonic 
atoms in an external trapping 
potential can be formulated within the non-relativistic 
quantum field theory [9-11]. 
The Lagrangian density of the Schr\"odinger field is given by 
$$ 
{\cal L} =  
{\hat \psi}^+({\bf r},t) \left[ i\hbar {\partial \over \partial t} 
+{\hbar^2 \over 2 m} \nabla^2 
- U({\bf r}) \right] {\hat \psi}({\bf r},t) 
$$
\beq 
-{1\over 2} \int d^3{\bf r}' 
\;\; {\hat \psi}^+({\bf r},t) 
{\hat \psi}^+({\bf r}',t) V(|{\bf r}-{\bf r}'|) 
{\hat \psi}({\bf r}',t) {\hat \psi}({\bf r},t) \; , 
\eeq 
where $U({\bf r})$ is the external potential and 
$V(|{\bf r}-{\bf r}'|)$ is the interatomic potential. 
The quantum bosonic field ${\hat \psi}({\bf r})$ 
satisfies the standard equal-time commutation rules.  
\par
The Lagrangian is invariant under the global $U(1)$ gauge 
transformation 
\beq
{\hat \psi} ({\bf r},t)\to e^{i\alpha} {\hat \psi} ({\bf r},t) \; ,
\eeq
which implies the conservation of the total number of particles 
\beq 
{\hat N}= \int d^3{\bf r} \;\; 
{\hat \psi}^+({\bf r},t){\hat \psi}({\bf r},t) \; .
\eeq 
The Bose-Einstein condensation (BEC) 
is the macroscopic occupation of the $N$-body ground-state 
$|{\bf O}>$ of the system. 
To study BEC a useful mean-field prescription is to separate out the 
condensate contribution to the bosonic field operator in the 
following way 
\beq 
{\hat \psi}({\bf r},t)= \phi ({\bf r},t) + 
{\hat \Sigma}({\bf r},t) \; , 
\eeq 
where $\phi ({\bf r},t)=
<{\bf O}|{\hat \psi}({\bf r},t)|{\bf O}>$ 
is the so-called macroscopic wavefunction (or order parameter) 
of the condensate, and ${\hat \Sigma}({\bf r},t)$ 
is the fluctuation operator, such that 
${\hat \Sigma}({\bf r},t)|{\bf O}>=0$. Note that this 
prescription breaks the $U(1)$ global gauge symmetry of the system [9-11]. 
\par 
Alkali vapors are quite dilute 
and at zero temperature the atoms are practically 
all in the condensate [4-6]. Thus 
we can neglect the quantum depletion due to the 
the operator ${\hat \Sigma}$ and the 
macroscopic wavefunction is normalized to the total number of atoms. 
Moreover, 
the range of the atom-atom interaction $V(r)$ is believed to be short in 
comparison to the typical length scale of variations of atomic 
wave functions. The atom-atom interaction is usually replaced by an effective 
zero-range pseudo-potential, 
$V(r)=g\delta^3(r)$, where $g={4\pi \hbar^2 a_s/m}$ 
is the scattering amplitude and $a_s$ is the s-wave scattering length. 
This scattering length is positive (repulsive 
interaction) for $^{87}$Rb and $^{23}$Na but negative (attractive 
interaction) for $^{7}Li$. 
Within these approximations, the Lagrangian density becomes a local 
function of the condensate wavefunction, namely 
\beq
{\cal L} = \phi^*({\bf r},t) \left[ i\hbar {\partial \over \partial t} 
+{\hbar^2 \over 2 m} \nabla^2 
- U({\bf r}) \right] \phi({\bf r},t) 
-{1\over 2} g |\phi ({\bf r}, t)|^4\; . 
\eeq 
By imposing the least action principle one obtains the following 
Euler-Lagrange equation 
\beq 
i\hbar {\partial \over \partial t} \phi({\bf r},t) 
= \left[ -{\hbar^2 \over 2 m} \nabla^2 + U({\bf r}) 
+ g |\phi({\bf r},t)|^2 \right] \phi({\bf r},t) \; ,  
\eeq 
which is a nonlinear Schr\"odinger equation and is called 
time-dependent Gross-Pitaevskii (GP) equation [7]. 
Note that the GP equation is nothing but the mean-field 
(Hartree) approximation of the exact time-dependent Schr\"odinger 
equation of the N-body problem, where the totally symmetric 
many-particle wavefunction $\Psi$ of the system 
is decomposed in the following way 
$\Psi({\bf r}_1,{\bf r}_2...,{\bf r}_N,t) = 
\phi({\bf r}_1,t) \phi({\bf r}_2,t) ... \phi({\bf r}_N,t)$. 

\section{Hydrodynamic equations of the Bose condensate} 
\par
The complex macroscopic wavefunction $\phi({\bf r},t)$ 
of the condensate can be written in terms of 
a modulus and a phase, as follows 
\beq
\phi({\bf r},t) = \sqrt{ \rho({\bf r},t)} \; e^{iS({\bf r},t)} \; . 
\eeq
The phase $S$ fixes the velocity field ${\bf v}=(\hbar /m)\nabla S$. 
The GP equation can hence be rewritten in the form of two coupled 
hydrodynamic equations [12] for the density and the velocity field 
\beq
{\partial \over \partial t} \rho + \nabla \cdot ({\bf v} \rho) = 0 
\eeq
\beq
m {\partial \over \partial t} {\bf v} + \nabla \left( U + g \rho - 
{\hbar^2 \over 2m \sqrt{\rho}}\nabla^2 \sqrt{\rho} + 
{mv^2\over 2} \right) =0 \; . 
\eeq
If the repulsive interaction among atoms is strong enough, then the 
density profiles become smooth and one can safely neglect the kinetic 
pressure term in $\hbar^2$ in the equation for the velocity field, 
which then takes the form 
\beq 
m {\partial \over \partial t} {\bf v} + \nabla \left( U + g \rho  + 
{mv^2\over 2} \right) =0 \; . 
\eeq 
\par
In the current experiments with alkali metal-atoms, 
the external trap is well approximated by a harmonic potential 
\beq
U({\bf r})={m\over 2} (\omega_1 x^2 + \omega_2 y^2 + \omega_3 z^2 ) \; . 
\eeq
The ground-state solution (${\bf v}=0$) of Eq. (10) is given by 
\beq
\rho({\bf r})  = g^{-1} \left[ \mu - U({\bf r}) \right] \; , 
\eeq
in the region where $\mu > U({\bf r})$, and $\rho = 0$ outside. The 
normalization condition on $\rho({\bf r})$ provides 
$\mu = (\hbar \omega_h/2)(15Na_s/a_h)^{2/5}$, 
where $\omega_h=(\omega_1 \omega_2 \omega_3)^{1/3}$ and 
$a_h=(\hbar/m\omega_h)^{1/2}$. 
\par
An analytic class of solutions [13] of the time-dependent 
hydrodynamic equations (8) and (10), which are valid when 
the condition $Na_s/a_h>>1$ is satisfied, 
is found by writing the density in the form 
\beq
\rho({\bf r},t) = a_0(t) - a_1(t)x^2 - a_2(t)y^2 -a_3(t)z^2 \; 
\eeq
in the region where $\rho({\bf r},t)$ is positive, and the 
velocity field as 
\beq
{\bf v}({\bf r},t) = {1\over 2}\nabla 
[ b_1(t) x^2 + b_2(t) y^2 + b_3(t) z^2 ] \; .
\eeq
The coefficient $a_0$ is fixed by the normalization of the 
density $a_0=(15N/8\pi)^{2/5}(a_xa_ya_z)^{1/5}$. 
By inserting these expressions in the hydrodynamic equations 
one finds 6 coupled differential equations for the time-dependent 
parameters $a_i(t)$ and $b_i(t)$. 
By introducing new variables $q_i$, defined by $a_i=m\omega_i^2 
(2gq_i^2 q_1 q_2 q_3)^{-1}$, the hydrodynamic equations give 
$a_i = {\dot q_i}/q_i$ and 
\beq
{\ddot q_i} + \omega_i^2 q_i = 
{\omega_i^2 \over q_i q_1 q_2 q_3} \; ,
\eeq
with $i=1,2,3$. The second and third terms give the effect of the external 
trap and of the interatomic forces, respectively. 
It is important to observe that, using the new variables $q_i$, the 
equations of motion do not depend on the value of the coupling 
constant $g$. In terms of $q_i$ the mean square radii of the condensate 
are $<x_i^2>=(2\mu/m\omega_i)q_i^2$ 
and the velocities are $<v_i^2>=(2\mu/m\omega_i){\dot q}_i^2$ [13]. 

\section{BEC Collective Modes and Chaos} 

The three differential equations (15) are the classical 
equations of motion of a system with coordinates $q_i$ and 
Lagrangian given by 
\beq
L={1\over 2} (\omega_1^{-2} {\dot q}_1^2 + \omega_2^{-2} 
{\dot q}_2^2 + \omega_3^{-2} {\dot q}_3^2) 
- {1\over 2}(q_1^2 + q_2^2 + q_3^2) - {1\over q_1 q_2 q_3} \; . 
\eeq 
This Lagrangian describes collective modes of the Bose condensate 
for $N a_s/a_h >>1$ [13]. As stressed previously, 
in such a case the collective dynamics of the condensate 
does not depend on the number of atoms 
or the scattering length. 
The minimum of the effective potential is at $q_i=1$, $i=1,2,3$. 
The mass matrix $M$ of the kinetic energy and 
the Hessian matrix $\Lambda$ of the potential energy 
at the equilibrium point are given by 
\beq 
M= 
\left( \begin{array}{ccc}  
\omega_1^{-2} & 0             & 0  \\  
     0        & \omega_2^{-2} & 0  \\
     0        & 0             & \omega_3^{-2}  \\ 
\end{array}  \right)
\;\;\;\;\;\;\; \hbox{and} \;\;\;\;\;\;\;
\Lambda = 
\left( \begin{array}{ccc} 
3 & 1 & 1  \\  
1 & 3 & 1  \\
1 & 1 & 3  \\ 
\end{array}  \right) \; .
\eeq 
The low-energy collective excitations of the condensate are 
the small oscillations of variables $q_i$'s around the equilibrium 
point. The calculation of the normal mode frequencies $\Omega$ 
for the motion of the condensate is reduced to 
the eigenvalue problem $\Lambda - \Omega^2 M =0$, which gives  
\beq 
\Omega^6
-3\left(\omega_1^2+\omega_2^2+\omega_3^2\right)\Omega^4+
8\left(\omega_1^2\; \omega_2^2+\omega_1^2\; \omega_3^2
+\omega_2^2\; \omega_3^2 \right)\Omega^2
-20\; \omega_1^2\; \omega_2^2\; \omega_3^2=0 \; .
\eeq   
Note that this formula has been recently obtained by using 
a variational approach [14] and also by studying 
hydrodynamic density fluctuations of the condensate [15]. 
For an axially symmetric trap, where 
$\omega_1=\omega_2=\omega_{\bot}$, the previous equation gives 
\beq 
\Omega_{1,2}=\left(2 + {3\over 2}\lambda^2 \pm 
{1\over 2}\left(16 + 9\lambda^4 - 16\lambda^2\right)^{1/2} \right)^{1/2} 
\; \omega_{\bot} \; ,
\;\;\;\;
\Omega_3 =\sqrt{2} \; \omega_{\bot} \; ,
\eeq 
where $\lambda = \omega_3/\omega_{\bot}$ is the asymmetry 
parameter of the trap [12]. 
Observe that the experimental results obtained 
on sodium vapors at MIT ($\lambda = \sqrt{8}$) 
are in good agreement with the theoretical 
values predicted by (19) [16]. 
In the case of an isotropic harmonic trap 
($\lambda = 1$) with frequency $\omega$, 
one obtains $\Omega_{1,2}=\sqrt{5}\; \omega$, 
$\Omega_3 = \sqrt{2} \; \omega$ [12]. 
\par 
In most experiments the confining trap has axial symmetry [4-6]. 
Let us analyze this case in detail. 
Because of the axial symmetry, we can impose $q_1=q_2=q_{\bot}$. 
Moreover, by using the adimensional time 
$\tau = \omega_{\bot}t$, the Hamiltonian of the 
BEC collective modes becomes 
\beq
H=p_{\bot}^2 + {1\over 2}\lambda^2 p_3^2 + q_{\bot}^2 
+ {1\over 2} q_3^2  + {1\over q_{\bot}^2 q_3} \; , 
\eeq
where $p_{\bot}=dq_{\bot}/d\tau$ and 
$p_3= \lambda^{-2}dq_3/d\tau$ are the conjugate momenta. 
Note that the condition $q_1=q_2$ restricts collective modes 
to monopole oscillations, where the third component of the 
angular momentum, that is a good quantum number, is zero [12]. 
Near the minimum of the potential the trajectories in 
the phase-space are periodic or quasi-periodic. On the contrary, 
far from the minimum, the effect of nonlinearity becomes important. 
As the KAM theorem [17] predicts, 
parts of phase space become filled with chaotic orbits, 
while in other parts the toroidal 
surfaces of the integrable system are deformed but not destroyed. 
\par
We use a symplectic Euler method (leap-Frog) to numerically compute 
the trajectories. The time-step is $\Delta \tau =10^{-4}$ and 
the energy is conserved to the sixth digit. The conservation 
of energy restricts any trajectory of the four-dimensional 
phase space to a three-dimensional energy shell. At a particular energy, 
the restriction $q_{\bot}=1$ defines a two-dimensional surface in 
the phase space, which is called Poincar\`e section. 
Each time a particular trajectory passes through the surface 
a point is plotted at the position of intersection $(q_3,p_3)$. 
We employ a first-order interpolation process to reduce inaccuracies 
due to the use of a finite step length. 

\begin{figure}[htb]
\centerline{\psfig{file=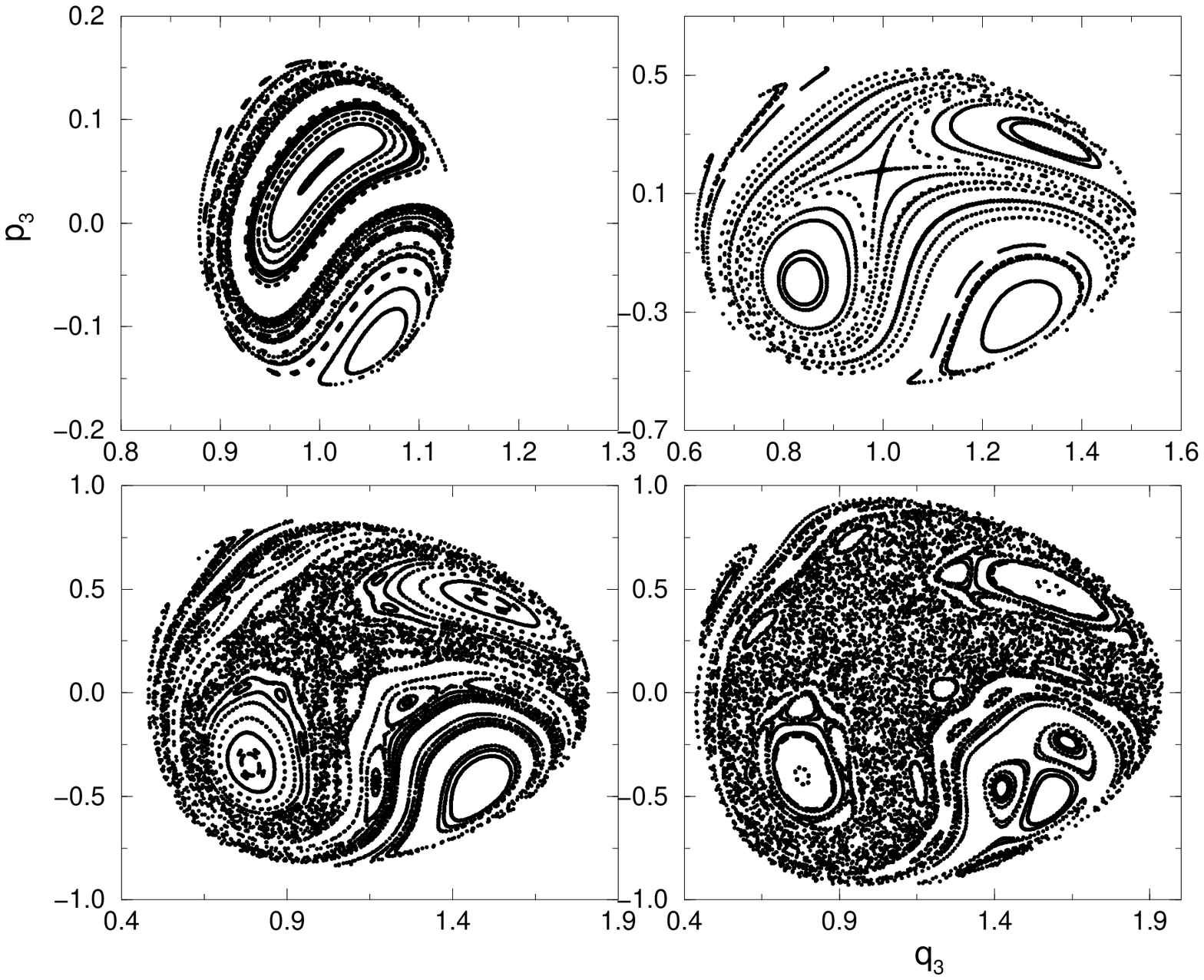}}
\end{figure}

{\bf Figure 1}: Poincar\`e sections with $\lambda=\sqrt{2}$. 
Each panel is at a fixed energy. 
From left to right and from top to bottom: $\chi = 1\%$, 
$\chi = 12\%$, $\chi = 28\%$, $\chi = 36\%$. 
$\chi$ is the relative increase of the energy with respect to the 
ground-state (minimum of the potential energy).

\par 
In Figure 1 we plot Poincar\`e sections of the system 
with $\lambda = \sqrt{2}$. In each panel there is a Poincar\`e 
section with a fixed value of the energy of the system. 
At each energy value, we have chosen different 
initial conditions [$q_{\bot}(0)$,$q_3(0)$,$p_{\bot}(0)$,$p_3(0)$] 
for the dynamics. Actually $p_{\bot}(0)$ has been fixed 
by the conservation of energy. Integration time is $400$ in 
adimensional units, that is less than $1$ second 
(the life-time of the condensate is about $10$ seconds). 
Note that the CPU time to calculate a Poincar\`e section 
with a dozen of initial conditions is about $30$ seconds. 
Chaotic regions on the Poincar\`e section 
are characterized by a set of randomly distributed points 
and regular regions by dotted or solid curves. 
Let $\chi$ be the relative increase of the energy with respect to the 
ground-state (minimum of the potential energy). 
For $\chi=1\%$ and $\chi=12\%$ the trajectories are still 
all regular but for $\chi = 28\%$ there is chaotic sea. 
For $\chi = 36 \%$ most trajectories are chaotic. 
\par
It is important to observe that 
a strong enhancement of nonlinear effects and eventually chaos 
can be obtained not only by increasing the energy 
but also by changing the anisotropy $\lambda$ of the trap. 
In fact, as shown in Ref. [13], for special values of $\lambda$, 
frequencies of different modes, or of their harmonics, can coincide. 
We plan to investigate in detail the onset of chaos for 
different configurations of the external trap. 

\section{Conclusions} 

In this short paper we have discussed the mean-field equations 
which describe the collective motion of a trapped weakly-interacting 
Bose condensate. We have shown by using Poincar\`e sections that 
for large energy values the system becomes chaotic. 
An important question is the following: Can BEC chaotic dynamics 
be experimentally detected? In our opinion the answer is positive. 
Nowadays non-destructive images of the dynamics of the condensate 
can be obtained. The radius of the condensate 
as a function of time can be detected and its power spectrum analyzed. 
In fact, one finds the the power spectrum of a chaotic signal 
is much more complex than for a regular one. Typically, one see few 
peaks in the regular signal and many peaks surrounded 
by a lot of noise in a chaotic one. 
Finally, we observe that various initial conditions for the collective 
dynamics of the condensate can be obtained by using laser beams or 
by modulating for a short period the magnetic fields which 
confine the condensate. 

\newpage 

\section*{References}

\begin{description}

\item{\ [1]} V.R. Manfredi and L. Salasnich, Int. J. Mod. Phys. 
E {\bf 4} 625 (1995).  

\item{\ [2]} V.R. Manfredi, M. Rosa-Clot, L. Salasnich, and S. Taddei, 
Int. J. Mod. Phys. E {\bf 5}, 519 (1996).  

\item{\ [3]} V.R. Manfredi and L. Salasnich, 
``Mean-Field and Nonlinear Dynamics in Many-Body Quantum Systems'', 
to appear in Proceedings of the 7th Workshop 
'Perspectives on Theoretical Nuclear Physics', 
Ed. A. Fabrocini {\it et al.} (Edizioni ETS, Pisa, 1999). 

\item{\ [4]} M.H. Anderson, J.R. Ensher, M.R. Matthews, C.E. Wieman, 
and E.A. Cornell, Science {\bf 269}, 189 (1995). 

\item{\ [5]} K.B. Davis, M.O. Mewes, M.R. Andrews, N.J. van Druten, 
D.S. Drufee, D.M. Stamper-Kurn, and W. Ketterle, Phys. Rev. Lett. {\bf 75}, 
3969 (1995).

\item{\ [6]} C.C. Bradley, C.A. Sackett, J.J. Tollet, and R.G. Hulet, 
Phys. Rev. Lett. {\bf 75}, 1687 (1995). 

\item{\ [7]} E.P. Gross, Nuovo Cimento {\bf 20}, 454 (1961); 
J. Math. Phys. {\bf 4}, 195 (1963); 
L.P. Pitaevskii, Zh. Eksp. Teor. Fiz. {\bf 40}, 646 (1961) 
[Sov. Phys. JETP {\bf 13}, 451 (1961)].

\item{\ [8]} D.S. Jin, J.R. Ensher, M.R. Matthews, 
C.E. Wieman, and E.A. Cornell, 
Phys. Rev. Lett. {\bf 77}, 420 (1996). 

\item{\ [9]} A.L. Fetter and J.D. Walecka, {\it Quantum Theory 
of Many-Particle Systems} (McGraw-Hill, New York, 1971). 

\item{\ [10]} K. Huang, {\it Statistical Mechanics} (Wiley, New York, 1963). 

\item{\ [11]} S.J. Chang, {\it Introduction to Quantum Field Theory}, 
Lecture Notes in Physics, vol. {\bf 29} (World Scientific, Singapore, 1990). 

\item{\ [12]} S. Stringari, Phys. Rev. Lett. {\bf 77}, 2360 (1996). 

\item{\ [13]} F. Dalfovo, C. Minniti, S. Stringari, and 
L. Pitaevskii, Phys. Lett. A {\bf 227}, 259 (1997); 
F. Dalfovo, C. Minniti, and L. Pitaevskii, 
Phys. Rev. A {\bf 56}, 4855 (1997). 

\item{\ [14]} E. Cerboneschi, R. Mannella, E. Arimondo, and 
L. Salasnich, Phys. Lett. A {\bf 249}, 245 (1998). 

\item{\ [15]} S. Stringari, Phys. Rev. A {\bf 58}, 2385 (1998). 

\item{\ [16]} D.M. Stamper-Kurn, H.J. Miesner, S. Inouye, 
M.R. Andrews, and W. Ketterle, 
Phys. Rev. Lett. {\bf }, 500 (1998).

\item{\ [17]} M.C. Gutzwiller, {\it Chaos in Classical and Quantum 
Mechanics} (Springer, New York, 1990); 
A.J. Lichtenberg and M.A. Lieberman, {\it Regular and Stochastic Motion}, 
(Springer, New York, 1983). 

\end{description}

\end{document}